# Analysing a built-in advantage in asymmetric darts contests using causal machine learning

Daniel Goller*

This version: August 2020

**Abstract:** We analyse a sequential contest with two players in darts where one of the contestants enjoys a technical advantage. Using methods from the causal machine learning literature, we analyse the built-in advantage, which is the first-mover having potentially more but never less moves. Our empirical findings suggest that the first-mover has an 8.6 percentage points higher probability to win the match induced by the technical advantage. Contestants with low performance measures and little experience have the highest built-in advantage. With regard to the fairness principle that contestants with equal abilities should have equal winning probabilities, this contest is ex-ante fair in the case of equal built-in advantages for both competitors and a randomized starting right. Nevertheless, the contest design produces unequal probabilities of winning for equally skilled contestants because of asymmetries in the built-in advantage associated with social pressure for contestants competing at home and away.

**Address for correspondence**: Daniel Goller,
Swiss Institute for Empirical Economic Research (SEW), University of St. Gallen, Varnbüelstrasse 14, CH-9000 St. Gallen, Switzerland, Daniel.Goller@unisg.ch.

---

* A previous version of the paper was presented at University of Göttingen, University of St. Gallen, and the YSEM, Zurich. Special thanks to the participants, as well as Robert Genthner, Sandro Heiniger, Michael Knaus, Alex Krumer, Michael Lechner, Gabriel Okasa, Anthony Strittmatter, and Michael Zimmert for helpful comments and suggestions. The usual disclaimer applies.

# 1 Introduction

Contest designers put a lot of effort in designing contests in various fields. For sports contests, fairness for the contestants and attractiveness for the spectators are the main restrictions (Szymanski, 2003; Wright, 2014; Arlegi and Dimitrov, 2020). The main fairness criterion for contests is fulfilled if equally skilled contestants have equal winning probabilities (e.g. Arlegi and Dimitrov, 2020). To ensure this, most contest designs balance potential advantages, e.g. home and away matches within one contest to balance the home advantage. Hence, many contest studies analyse symmetric contests that are designed to have either a balanced or no technical advantage but empirically uncover side effects, such as the order of actions (Ginsberg and Van Ours, 2003) or scheduling effects (Goller and Krumer, 2020).

In contrast, we investigate the fairness of an unbalanced, one match sequential contest with two competitors in the sport of darts. This contest has a built-in advantage (BIA) for the first moving contestant, who has an advantage in potentially more, but never less situations within a match.[1] In this study, we analyse a simple question: Is the two player, sequential contest with BIA a fair contest, or are there specific groups of individuals who are systematically disadvantaged by the contest design?

The outcome of a contest is about human behaviour, and even though behavioural responses are potentially among the most individual, there is hardly any personalized evidence. The importance of individual behavioural responses calling for more individualized effects estimation is demonstrated in the work of González-Díaz, Gossner, and Rogers (2012). They found large differences of individual 'critical abilities' of tennis players affecting the performance and success. While it is impossible to observe or estimate a true individual effect owing to the unobservable outcomes in situations of non-realized interventions, there is a huge potential to find or describe those individuals who benefit most or least from some given treatment, which is the BIA in the contest design. For designing rules for competitions, identifying the most benefiting group of individuals might help improve contest design in terms of fairness, attractiveness, or competitive balance. This is especially true if specific groups of individuals systematically realize differential effects and are therefore (dis-)advantaged by the contest design.

---

[1] Sequential contests with different kinds of advantages are widely present. Examples are legislators giving an advantage, in form of patents, to a first moving contestant in a market, job posting committees giving an advantage to specific groups of the population in a contest for a job, incumbency advantage in political elections, or in sports contests such as the snooker championship or NBA and NHL playoffs.

The relation of incentives and performance is widely discussed in the literature and is among the fundamentals in contests (Baumeister, 1984; Rosen; 1986; Ehrenberg and Bognanno, 1990; Ariely, Gneezy, Loewenstein, and Mazar, 2009). Humans behave differently in situations with increased level of incentives, which are present in many fields of daily life; such as job talks, presentations or speeches in the public, competitions and decisions at the workplace or in sports. Often those situations are subject to higher rewards or reputation that might increase or decrease the performance of individuals. In darts, social incentives might play a crucial role, owing to euphoric spectators and the otherwise very standardized environment. For athletes competing near their hometown the stakes are higher. Having a supporting crowd around might be motivating, which we call social support from now on. Social pressure is the counterpart, as being watched can also be a burden inducing pressure.[2] The increased focus magnifies the reputation gained if performing well, as well as the loss of reputation, otherwise. This is a known phenomenon in the social facilitation literature (e.g. Zajonc, 1965; Butler and Baumeister, 1998), which is found to have an influence even in cases in which not the athlete, but only the audience has high expectations (Baumeister, Hamilton, and Tice, 1985; Strauss, 1997).

---

[2] While social support is supposed to improve performance, social pressure results in worse performance (Zajonc, 1965; Baumeister, Hamilton, and Tice, 1985).

This motivated us to analyse differential effects in the BIA on an individualized level, as well as of the respective groups associated with social incentives and ability in a flexible way. To the best of our knowledge we are the first to empirically analyse a BIA in an asymmetric two players sequential contest. This study especially contributes in investigating the fairness of such a contest with a thorough analysis of the BIA using state-of-the-art methodology.

This work relies on recent advances in econometric methods made in the emerging *Causal Machine Learning* literature. These new methods are more flexible compared to parametric approaches. Moreover, they can deal flexibly with rich data, which makes some identifying assumptions more credible, and they are useful for the analysis of heterogeneous treatment effects.

For this, we use a dataset from the increasingly popular sports of darts, which offers several benefits when investigating individual responses. Particular for darts is a friendly but euphoric atmosphere created by the spectators. Moreover, conditions at different venues are highly standardized. A noteworthy advantage in comparison to various other sports is that there is no direct interaction between the contestants, which could lead to problems in assessing individual performance, otherwise. Furthermore, the outcome is precisely measurable, the rewards are high enough for competitors to take that task seriously, humans are observed while performing their usual job (Levitt and List, 2008), clear rules avoid subjective decisions by referees, and almost no external influences.

As expected, we found the average effect for the technical advantage of moving first to deliver an about 8.6 percentage points higher probability to win a match. Despite this advantage, the contest is ex-ante fair for players with symmetric BIAs and a randomized allocation to be the first-mover. The contest cannot be regarded as fair in case of systematically differential BIAs, which is systematic effect heterogeneity associated to specific groups. The empirical analysis shows equal BIAs for equally skilled contestants, but the personalized effects span from about -5 to +15 percentage points for individuals with different characteristics showing that there is a substantial heterogeneity in the effect of the BIA. Those contestants with lower performance measures and less experience profit most from the BIA. Differential effects are found in line with the social pressure hypothesis, as contestants playing in a neutral environment benefit from an 8.8 percentage points higher treatment effect compared to those playing in a supportive environment. Interestingly, this differential effect in the BIA is found only for the first-mover player, while the second moving player is not affected by this social pressure. Therefore, this is not a general home disadvantage but is related to social pressure only for the first moving contestant. This results in an unfair contest design, since equally skilled contestants do not necessarily benefit from equal winning probabilities.

In the following section the related literature is discussed. Section 3 provides an insight into the setting of darts and details the data used. Section 4 presents the methodological challenges and the estimation procedures used to obtain the results in Section 5, followed by a conclusion in Section 6.

## 2    Literature review

There are few studies on the sport of darts, starting with Tibshirani, Price, and Taylor (2011) analysing the statistics behind darts and the ways to get the highest expected payoffs. Liebscher and Kirschstein (2017) predicted winning probabilities for individual players in the world darts championship. While they implicitly incorporated the BIA in their prediction, they did not intend to estimate causal effects. Ötting, Deutscher, Langrock, Gehrmann, Schneemann, and Scholten (2020), and Klein Teeselink, van Loon, van den Assem, and van Dolder (2020) discovered no or low performance decrements under pressure for professional darts players by looking at more or less pressure owing to varying importance of situations within a match. Substantial decrements among youth and amateur players are found by Klein Teeselink et al.

(2020), which might suggest a selection of more choking resistant individuals among professionals.

One of the most fundamental relations in contests is the role of incentives on performance, discussed in the behavioural and psychological (e.g. Baumeister, 1984; Masters, 1992; Butler and Baumeister, 1998; Ariely et al., 2009) and the economics literature (e.g. Stiglitz, 1976; Shapiro and Stiglitz, 1984; Rosen, 1986; Prendergast, 1999; Lazear, 2000).[3] Several works have investigated the role of social incentives, for example Cao, Price, and Stone (2011) and Toma (2017), which used data on basketball and found no effect of pressure on performance when a team plays at home compared to road games. Other works find this to be a source of pressure that impacts performance negatively. Ariely et al. (2009) found high social rewards to act counterproductive for performance in a set of experiments. Harb-Wu and Krumer (2019) investigated shooting performance in biathlon and found home athletes to miss more shots compared to athletes from other countries. Furthermore, the authors suggested that performance decrements are present only for the best ranked quarter of athletes. Despite the direct psychological effect, there is an indirect effect from public expectations (Baumeister, Hamilton, and Tice, 1985, Butler and Baumeister, 1998, Strauss, 1997). Furthermore, Baumeister et al. (1985) and Strauss (1997) found performance decrements among individuals in cases in which the audience, but not the individual expects success.

Symmetric contests with either a balanced or no advantage are analysed with regard to differential effects for specific groups in several works. Ginsburg and Van Ours (2003) found different winning probabilities in a musical contest depending on the order of actions. Dohmen (2008) and Harb-Wu and Krumer (2019) reported performance decrements among home contestants compared to non-home contestants. In other works, the groups of first-movers (e.g.

---

[3] While there is a general agreement, that a small increase in incentives has a positive effect on performance, economic theory, such as the efficient wage hypothesis, suggests that stronger incentives lead to more effort and higher output. Behavioural research finds performance decrements under circumstances of increased importance mainly for skill-based tasks. As it can be assumed that darts is a skill-based task, we focus on research from this literature.

Apesteguia and Palacios-Huerta, 2010) or the second-movers (e.g. Page and Page, 2007) are found to have higher winning probabilities than the respective other. Recently, Goller and Krumer (2020) established differential winning probabilities in home matches for underdog teams depending on scheduling of the matches. Asymmetric contests, in which certain contestants are favoured by the contest design, are examined in various theoretical works. Examples are the role of the incumbency advantage for autocrats' investment decisions (Konrad, 2002) or in political campaigns (Meirowitz, 2008), and more generally the influences of head-starts or handicaps in contests (Kirkegaard, 2012; Segev and Sela, 2014). In a more general scheme, a BIA is found for relative age effects in youth sports, in which children born at a certain time of year are favoured by the calendar year system, which groups children for competitive purposes (for a review, see Musch and Grondin, 2001; for a solution on how operational research can improve fairness, see Hurley, 2009).

Using sports data, machine learning methods are up to now almost exclusively used for prediction tasks, rarely in causal studies for average effects but, to the best of our knowledge, not yet for the systematic estimation of heterogeneous effects.[4] Especially in recent years, a number of methods have been proposed in the *Causal Machine Learning* literature for estimating effect heterogeneity (Athey, Tibshirani, and Wager, 2018; Wager and Athey, 2018; Lechner, 2018; Zimmert and Lechner, 2019; among others). Machine learning methods (or statistical learning methods, see e.g. Hastie, Tibshirani, and Friedman, 2009) were implemented or modified to suit the classical causal framework and developed to become useful in analysing causal questions (Athey, 2017; Athey and Imbens, 2019). While there is some work establishing theoretical guarantees (e.g. Chernozhukov et al., 2018; Wager and Athey, 2018; Zimmert and

---

[4] Examples of prediction tasks are forecasting the winner of the darts world championship (Liebscher and Kirschstein, 2017) or predicting the final table in the German soccer Bundesliga (Goller et al., 2018). Average causal effects estimations can be found for schedule-related issues in soccer, as in Krumer and Lechner (2018) and Goller and Krumer (2020). Further, Faltings, Krumer, and Lechner (2019) used a method from the causal machine learning literature in their work on favouritism in the Swiss soccer league but did not find any systematic heterogeneity in their effects.

Lechner, 2019) and simulating the performance of the newly developed estimators (notably Knaus, Lechner, and Strittmatter, 2020a), empirical applications of those new estimators and the systematic estimation of heterogeneous effects become increasingly popular (examples are Davis and Heller, 2017; Cockx, Lechner, and Bollens, 2019; Athey and Wager, 2019; Knaus, Lechner, and Strittmatter, 2020b).

# 3 Setting and data

## 3.1 Setting

Darts players commit themselves to play in one of the two major federations, the British Darts Organization (BDO) or the Professional Darts Corporation (PDC). While both federations hold their own tournaments, the PDC receives more media attention and distributes higher prize money than the BDO.[5] The most important tournaments are the (PDC/BDO) World Darts Championship held in December and January each year. Additionally, during the year, several other major tournaments are held in different places. The majors are joined by different minor tournaments throughout the year, relevant for accumulating ranking relevant prize money and qualification for the major tournaments. For a full list of tournaments considered in this work, see Appendix E.

All BDO and PDC darts tournaments operate under the rules of the Darts Regulation Authority. Most matches are played in the (best-of-K) legs format, where K is an odd number. This implies that a player who wins (K+1)/2 legs wins the match. To win a leg, the contestants must score exactly 501 points and complete with a 'double' – a special region on the border of the darts board in which the score achieved is doubled. The two contestants perform their moves sequentially; for each move, three darts are to be thrown. Once a player reaches 501 points, the opponent is not allowed to 'catch-up'. Therefore, the first-mover in a given leg potentially has up to three darts more compared to the non-starting player. Having an odd number of maximum legs played, the first-mover has a technical advantage for the whole match.

---
[5] There are also combined tournaments for which the players of BDO and PDC can participate. Further, the players may change their affiliation any time.

Before each match the starter is determined in a shootout, which is one dart each player, with the player closest to the centre of the darts board starting the match. How this challenge of non-random determination of moving first is solved is discussed in Section 4.2. For a more formal model the interested reader can refer to Appendix C.

## 3.2 Data base

To investigate the BIA, we use data from the sports of darts. Starting from the year 2009 until 2019, matches from the most important BDO and PDC tournaments were extracted from the software, *dartsforwindows*, containing information on match statistics and outcomes. This resulted in a total of 11,604 matches played by 394 different players.[6] Importantly, the variable of interest, i.e. *starting the first leg*, as well as the outcome variable, *winning the match*, is generated. This is complemented by venue and tournament characteristics, such as prize money, which is standardized to make it comparable over the studied time period.[7] Personal characteristics of the players, such as nationality, hometown, and date of birth, among others are collected and utilized to create additional variables. Those contain the *age* or the number of years the player has played darts at the time of the respective match, or the distance between the hometown and the venue. *Home* and *Venue in country of birth* variables are created if the player lives within 100 kilometres of the venue, and is born in that country, respectively.[8] Since matches vary in their potential length, the data base contains the maximum number of legs to play (*bestoflegs*) in the respective match. Finally, (pre-match) performance measures and players statistics, such as the *3-darts average* (cumulated 3 darts score, averaged over all the matches in the past 2 years), *rankings*, etc., are added. For the full list of covariates and sources, the reader can refer to Appendices A and F, respectively.

---

[6] After excluding all matches that included players with less than 5 matches, and few other players whose characteristics were missing.

[7] Since prize money in tournament *m* raises substantially over the years *t* this is standardized as:
$Prizemoney\ standardized = \frac{Prizemoney_{mt} - \min_t(Prizemoney_t)}{\max_t(Prizemoney_t) - \min_t(Prizemoney_t)}$.

[8] The main definition of the variable *Home* of a radius of 100 kilometres around the venue is chosen to specify the hometown as close to the venue as possible, i.e. at a distance that family and supporters can easily accommodate, and to observe enough individuals as 'home contestants' to have sufficient statistical power. Further, to investigate the robustness regarding this decision the radius is varied to 50, 150, and 200 km around the venue.

## 3.3 Descriptive statistics

*Table 1: Descriptive Statistics*

| Variable | (1) Mean | (2) Starting Player | (3) Non-starting Player | (4) Diff / SD |
|---|---|---|---|---|
| Game Outcomes | | | | |
| Wins Match | 0.50 | 0.55 | 0.45 | 0.10 |
| Wins First Leg | 0.50 | 0.63 | 0.37 | 0.27 |
| Game Characteristics | | | | |
| Starts First Leg | 0.50 | 1.00 | 0.00 | |
| Televised | 0.32 | | | |
| Ranking Tournament | 0.90 | | | |
| Prize Money | 51145 (79918) | | | |
| Prize Money standardized | 0.13 (0.25) | | | |
| Player Characteristics | | | | |
| 3 Darts Average | 91.82 (4.53) | 91.99 (4.49) | 91.65 (4.56) | 7.54 |
| Ranking Last Year | 121.74 (292.77) | 113.01 (254.99) | 130.47 (326.21) | 5.97 |
| Accumulated Matches | 231.15 (270.17) | 238.13 (274.46) | 224.17 (265.81) | 5.17 |
| Left handed | 0.07 | 0.07 | 0.07 | 0.80 |
| Years Playing | 19.97 (9.82) | 20.07 (9.79) | 19.87 (9.85) | 2.05 |
| Years Playing Professional | 11.91 (6.21) | 11.90 (6.17) | 11.91 (6.25) | 0.23 |
| Age | 36.54 (9.73) | 36.46 (9.63) | 36.61 (9.83) | 1.57 |
| Venue in country of birth | 0.38 | 0.38 | 0.38 | 0.21 |
| Home | 0.08 | 0.08 | 0.08 | 0.36 |
| Distance to Venue | 1410.25 (3271.76) | 1422.84 (3307.26) | 1397.66 (3241.93) | 0.77 |
| Observations | 11604 | 11604 | 11604 | |

Notes: This table presents average values and standard deviations (in parentheses for non-binary variables) of selected variables. In the last column the difference for the outcome variables from starting (2) and non-starting (3) player, respectively the standardized difference for the player characteristics is reported. Each observation represents one match. Values that are equal for all columns are reported only once.

The outcome variable under consideration is *winning the match* and is depicted in the descriptive statistics presented in Table 1. Further, information is provided by categorising the players according to the variable of interest, i.e. *starting the first leg*, in columns (2) and (3).

The difference in the outcome variables for starting and non-starting players can be observed in column (4). If starting the first leg is determined randomly (instead of performing the shootout) these numbers would be the average treatment effects. Descriptive evidence that moving first is subject to selection effects can be observed in the standardized differences (SD) reported in column (4) for the players characteristics.[9] Starting players have on an average a higher *3-darts average*, a better *ranking* position, and more *accumulated matches*. For the other characteristics the standardized difference is low. The full list of variables can be found in Table A.1 in Appendix A.

## 4  Methodology

### 4.1  Notation and framework

The typical notation for binary treatment effects estimation, following Rubin (1974) is used. Suppose the outcome obeys the observational rule: $Y_i = D_i Y_i(1) + (1 - D_i) Y_i(0)$, where $D_i$ denotes the treatment status $d \in (0,1)$, $Y_i(d)$ the potential outcome under treatment status $d$. Furthermore, we define $X_i$ to contain the covariates necessary to account for confounding, and $Z_i$ represents those variables investigated in the heterogeneity analysis.[10]

The first estimand of interest is the average treatment effect (ATE), $\theta = E(Y_i(1) - Y_i(0))$. This represents the average effect for all units on the highest level of aggregation. Contrarily, the estimand on the lowest aggregation level is the individualized average treatment effect (IATE), $\theta(x) = E(Y_i(1) - Y_i(0)|X_i = x)$. The group average treatment effect (GATE) represents an intermediate aggregation level according to heterogeneity variables $Z_i$, $\theta(z) =$ $E(Y_i(1) - Y_i(0)|Z_i = z)$. Both conditional average treatment effects (CATEs), the GATEs and IATEs, are useful for detecting heterogeneity among the observed units, which are 'hidden' in ATE estimates.[11]

---

[9] It is unclear in the literature how large the standardized difference can be to still claim the average of the value of a respective variable to be balanced among treated and non-treated. Moreover, this is only an indication for average balancing. In this work we were very cautious even with seemingly low SD and took non-random treatment allocation seriously, as described in Section 4.2.

[10] In principle, $X_i$ and $Z_i$ might contain distinct variables or overlap, partly or completely. In this work $Z_i$ is an ad-hoc selected subset of $X_i$, used for the heterogeneity analyses and investigating the hypotheses. Throughout the work, random variables are indicated by capital letters and realizations of these random variables by lowercase letters.

To understand the relationship of those three estimands of interest, integrating the IATEs over the characteristics of the groups $Z_i = z$ leads to the GATEs, while integrating over the characteristics of the entire population results in the ATE. Finally, only one of the potential outcomes is observable, as a unit can either be treated ($D_i = 1$) or non-treated ($D_i = 0$); the other remains counterfactual. In the following section, we shall discuss how to 'solve' this fundamental problem of causal inference (Holland, 1986).

## 4.2 Identification

Most crucial for estimating causal effects is a credible identification strategy. In the case of randomized treatment assignment, non-treated units can directly be used to construct the counterfactual outcome. In the underlying case, with non-random treatment assignment, we impose the following assumptions to identify the estimands of interest based on selection-on-observables. First, the conditional independence assumption (CIA): $Y_i(1), Y_i(0) \perp D_i|X_i = x$, and second, common support (CS): $0 < P[D_i = 1|X_i = x] < 1$.

The first assumption, the CIA, states that the potential outcomes are independent of the treatment assignment conditional on the confounders. This implies that all variables affecting both, treatment assignment and outcome, are observed and contained in $X_i$. The second assumption, CS, requires that treatment possibilities are bounded away from 0 and 1. This assumption is testable and of no issue in this work (see Figure A.1 in Appendix B.1).

At this point, we would like to highlight the two different roles of covariates. First, covariates (or confounders; $X_i$) are required to make the CIA credible. In other words, the characteristics that are responsible for selection into the treatment status have to be accounted for to obtain an 'as-good-as-randomized' situation. This is necessary to obtain causal treatment effects, which are free from selection bias. Second, covariates (or heterogeneity variables; $Z_i$) are used to form groups of observations for which heterogeneous effects are to be estimated.

---

[11] GATE and IATE are special cases of the CATE. The more precise terms are used in this section; throughout this work, the terms are used synonymously.

In this application, the non-random determination of the treatment, i.e. starting the first leg, is a result of the shootout to select the starter as already discussed in Section 3.1. The consequence of this shootout is arguably mainly influenced by the ability of the players and their experience with this situation. As indicated in Section 3.2, there are two measures of ability, the previous ranking and the performance in form of the previous average scores achieved. Experience is measured by the cumulated number of matches played, age, number of years of playing darts and the number of years played at a professional level. Furthermore, we control for other variables potentially influencing the selection into treatment as described in Section 3. Since it is unclear in which (functional) form the potential confounders are most relevant to account for the selection into treatment, a flexible estimation technique, namely, a double machine learning approach with non-parametrically estimated nuisance functions, described in the following section, will be used. Having this set of observed potential confounders, it is credible that the CIA is satisfied.

## 4.3 Methods

If one considers using a classical linear regression approach to estimate a treatment effect, there are some implicit assumptions to think about. As already discussed, the conditional independence assumption is crucial. Further, there are the assumptions of a constant treatment effect and a linear effect of the confounders $X_i$ on $Y_i$. In most studies, there are neither scientific nor methodological reasons to motivate these latter assumptions. Therefore, adopting a methodology not imposing those assumptions is desirable. The upcoming *Causal Machine Learning* literature is agnostic about these assumptions.

The growing literature offers several solutions to estimate treatment effects in a flexible way. For a summary and simulation study covering many of the methods, refer to Knaus, Lechner, and Strittmatter (2020a). They found four causal machine learning methods with good performance in all their simulated settings, while other methods were unstable or did not perform well for estimating the CATE and ATE. Specifically, those are LASSO with covariate modification and efficiency augmentation (Tian et al., 2014), LASSO with R-Learning (Nie and Wager, 2017), Causal Forest with local centring (Athey, Tibshirani, and Wager, 2019; Lechner, 2018), and Random Forest with Double Machine Learning (Chernozhukov et al., 2018). For this work we chose to use Double Machine Learning, for which we have theoretical guarantees required for the strategy of our application, e.g. for the ATE (Chernozhukov et al., 2018) or the CATE (Semenova and Chernozhukov, 2017; Zimmert and Lechner, 2019). To check the sensitivity of the results being method dependent we chose to additionally perform the analysis using a Modified Causal Forest (Lechner, 2018).[12]

In the following sections, we build on the observation of the two different roles of covariates. First, the procedure to account for selection into treatment by controlling for (potential) confounding factors is introduced in Section 4.3.1. Second, two ways how heterogeneity variables are used to investigate granular treatment effects are described in Section 4.3.2.

**4.3.1 Double machine learning**

The first stage of the estimation procedure uses the so-called *Double Machine Learning* (DML), introduced by Chernozhukov et al. (2018). This approach to overcome the selection problem builds on the augmented inverse probability weighting procedure, going back to Robins, Rotnitzky, and Zhao (1994, 1995):

$$Y_i^* = \mu_1(X_i) - \mu_0(X_i) + \frac{D_i(Y_i - \mu_1(X_i))}{p(X_i)} - \frac{(1-D_i)(Y_i - \mu_0(X_i))}{1-p(X_i)},$$

and involves three nuisance parameters. $\mu_1(X_i) = E(Y_i|D_i = 1, X_i)$, modelling the conditional outcome mean if treated, $\mu_0(X_i) = E(Y_i|D_i = 0, X_i)$, the conditional outcome mean if not treated and $p(X_i) = E(D_i|X_i)$, the conditional probability to be treated.[13] This results in the orthogonal score ($Y_i^*$), which has various applications. The expected value of the orthogonal scores, i.e. $E(Y_i^*)$, is the average treatment effect. The lowest level of aggregation is obtained as $E(Y_i^*|X_i = x)$, while group average treatment effects can be obtained by the expected value conditional on the groups defined in $Z$ as $E(Y_i^*|Z_i = z)$, which are discussed in detail in Section 4.3.2. The three nuisance parameters in general can be estimated by any well-suited estimation technique.[14]

---

[12] An advantage is the easy implementation of cluster-based sampling, which is helpful in case observations are clustered. The potential drawback of the Modified Causal Forest method is efficiency loss owing to the requirement of sample splitting for weight-based inference, which can be circumvented by cross-fitting in the Double Machine Learning approach. Moreover, it works with only discrete or discretized heterogeneity variables. For a more detailed description the interested reader is referred to Lechner (2018) in general, as well as the appendix for a short introduction and the process of its implementation in this work.

To overcome the linearity assumption of effects by the confounders on the outcome, the non-linear and non-parametric Random Forest is used to estimate the nuisance parameters. The Random Forest algorithm, developed in Breiman (2001), is built as an ensemble of single Regression Trees, which are to some extent randomly constructed.[15] Each Regression Tree recursively splits the space of covariates into non-overlapping areas to minimize the MSE of the outcome prediction until it reaches some stopping criteria. The resulting structure is reminiscent of a rotated tree; one can observe the trunk gradually splitting up into finer branches. The averages of the outcomes of those observations falling into the same end-nodes (so called leaves) provide the predictions of the tree. Combining several of those tree predictions results in the final predictions of the Random Forest.[16] This DML step is to remove any (potential) confounding issues. The resulting orthogonal score is free from selection effects, and treatment effects can be constructed at various levels of aggregation.[17]

---

[13] Following Chernozhukov et al. (2018), the nuisance parameters are estimated based on a 2-fold cross-fitting, in which first the sample is randomly divided into 2 equally sized folds. In each of the folds, the prediction model is estimated, while in the respective hold-out fold, the model is used to predict the nuisance parameters. While no observation is used to predict its own nuisance, we can use all observations without running into problems of overfitting.

[14] There are requirements on the rate of convergence of the estimators (depending on level of aggregation, cross-fitting, etc.). Those requirements are met for standard regression models, such as OLS and Probit, as well as for specific methods coming from the machine learning literature, such as Random Forest or LASSO and others. In this work, the Random Forest is used, for which the required rates are given (compare Wager and Walther, 2015).

[15] For each tree 50% of the data are subsampled. To further de-correlate the trees at each split, only a subset of the available covariates is randomly chosen.

### 4.3.2 Conditional average treatment effects

As already mentioned, the scores can be used to directly obtain the average treatment effect by $\theta = E(Y_i^*)$. To estimate effects beneath the average level, the obtained orthogonal scores from the DML procedure can be used in different ways. Consider first using the Best Linear Predictor (BLP) framework proposed in Semenova and Chernozhukov (2017). Here, the orthogonal scores are used as pseudo-outcome in an ordinary least squares regression on covariates to solve the minimization problem: $\hat{\beta} = argmin \sum_{i=1}^{N}(\widehat{Y_i^*} - \tilde{x}_i\beta)^2$, with $\tilde{x}_i$ containing a constant and $x_i$. The fitted values, $\hat{\theta}(\tilde{x}_i) = \tilde{x}_i\hat{\beta}$ are the best linear predictors of the IATEs. In fact, $\tilde{x}_i$ can be replaced by any subset of the covariate space. For example, replacing it by only a constant leads to the ATE; replacing $\tilde{x}_i$ by $\tilde{z}_i = (constant, z_i)$ leads to the GATEs. Standard errors can be computed as heteroscedasticity robust standard errors, which are valid, as shown in Semenova and Chernozhukov (2017).[18] Once the IATEs are estimated, they can be analysed in various ways. One possibility is to evaluate the distribution of the effects, to observe how different the effects are for the range of observations. Chernozhukov, Fernández-Val, and Luo (2018) suggested the Sorted Effects method, which sorts the individualized effects according to the effect size, coming with valid confidence intervals. Furthermore, this can be used to analyse whether the most and least affected individuals differ according to their characteristics.

---

[16] All the necessary tuning parameters are determined in a data-driven way using cross-validation. For more details on Random Forests, the interested reader can refer to the initial study by Breiman (2001) or the book of Hastie, Tibshirani, and Friedman (2009).
[17] For a more detailed discussion, the interested reader can refer to Knaus (2020).
[18] In a sensitivity check, we account for potential clustering on the individual contestant level.

The discussed method has its strength in analysing the total range of individualized effects and compactly summarising heterogeneity. The drawback is that an assumption of linearity of effects is imposed, as one runs quickly into the 'curse of dimensionality' with more than few heterogeneity variables when using non-parametric methods. To investigate specific hypotheses, involving few dimensions of the covariate space, Zimmert and Lechner (2019), as well as Fan, Hsu, Lieli, and Zhang (2019), proposed a non-parametric approach to not be dependent on the linearity assumption of the best linear predictor. The already estimated orthogonal scores can be used with (few) selected heterogeneity variables in a classical non-parametric kernel regression as follows:

$$\hat{\theta}(z) = \sum_{i=1}^{N} \frac{\mathcal{K}_h(z_i - z)\widehat{Y_i^*}}{\sum_{i=1}^{N} \mathcal{K}_h(z_i - z)}$$

With $\mathcal{K}_h$ being a kernel function with bandwidth $h$, determined by cross validation and 90% undersmoothing, as suggested in Zimmert and Lechner (2019). The drawback of this procedure is that the dimension of $Z_i$ is limited to obtain the required asymptotic guarantees. $Z_i$, in this case, includes one or two heterogeneity variables, which is unproblematic from a theoretical perspective and sufficient to analyse the hypotheses of interest in this work.

## 5 Results

### 5.1 Average treatment effect

The first result of the analysis is the average effect of the BIA for the starting player on winning the match. Column (1) in Table 2 shows the average treatment effect. The effect of the technical advantage with 0.0865 is large and precise enough to be statistically different from zero. The average effect for the starting contestants, therefore, amounts to 8.65 percentage points higher winning probability for the match. This effect is in line with what we would expect owing to the technical advantage implicit in the contest design.

*Table 2: Base results - best linear predictors*

|  | (1) ATE | (2) Home | (3) Country of birth | (4) All |
|---|---|---|---|---|
| Intercept | 0.0865*** | 0.0939*** | 0.0853*** | 0.0143 |
|  | (0.0089) | (0.0093) | (0.0112) | (0.2406) |
| Home |  | -0.0885*** |  | -0.0987*** |
|  |  | (0.0323) |  | (0.0347) |
| Venue in country of birth |  |  | 0.0032 | 0.0248 |
|  |  |  | (0.0184) | (0.0199) |
| Prize money standardized |  |  |  | -0.0736* |
|  |  |  |  | (0.0488) |
| Ranking |  |  |  | 0.0000 |
|  |  |  |  | (0.0000) |
| 3-darts average |  |  |  | 0.0009 |
|  |  |  |  | (0.0026) |
| Acc. matches |  |  |  | -0.0001 |
|  |  |  |  | (0.0000) |
| Years playing |  |  |  | -0.0016 |
|  |  |  |  | (0.0016) |
| Years prof. |  |  |  | -0.0026 |
|  |  |  |  | (0.0024) |
| Age |  |  |  | 0.0020 |
|  |  |  |  | (0.0016) |
| Best of legs |  |  |  | 0.0011 |
|  |  |  |  | (0.0020) |
| Observations | 11604 | 11604 | 11604 | 11604 |

Notes: Best linear prediction as described in Semenova and Chernozhukov (2017); heteroscedasticity robust standard error in parentheses. *Home* is defined as the hometown of the contestant being in the radius of 100km to the venue. *, **, and *** represent statistical significance at the 10, 5, and 1% level, respectively.

While this is a sizeable effect, it would be interesting to note the kind of players that may profit more or less from this advantage. To analyse this, we look into more granular effects in the following sections.

## 5.2 Individualized average effects

Starting on the most granular level, we see in Figure 1 the individualized average treatment effects for the BIA sorted in size. The solid black line represents the average treatment effect with the dotted black lines being its confidence intervals. The solid blue line represents the sorted individualized effects, accompanied by the shaded blue confidence intervals.

*Figure 1: Sorted Effects*

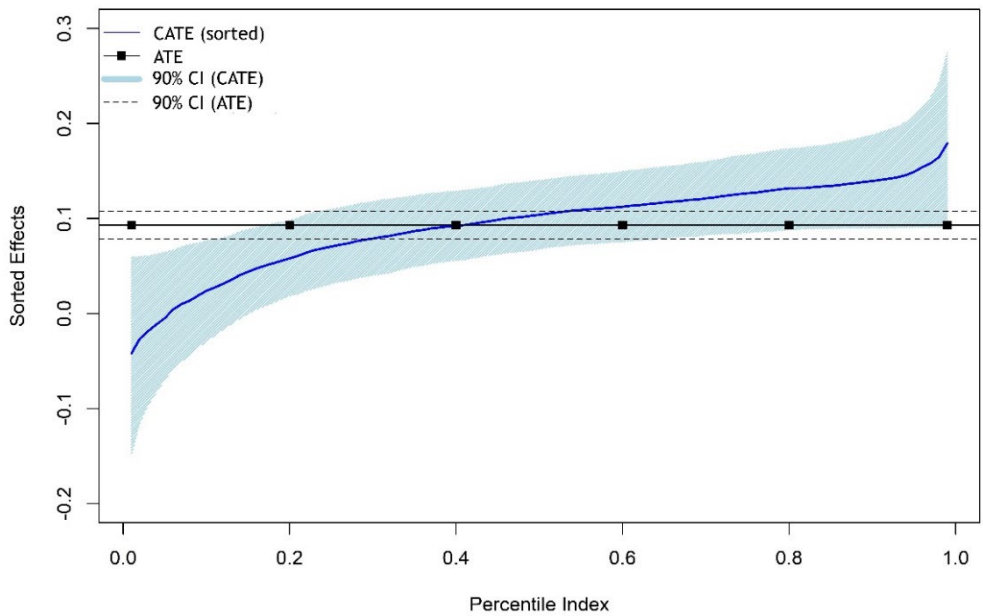

Notes: Sorted Effects. 999 (weighted) bootstrap replications. Bias corrected. The blue line represents the sorted Conditional Average Treatment Effects accompanied by the shaded 90% confidence interval. The black line with squares represents the ATE, accompanied by the dashed line representing the 90% confidence interval.

While most of the players have positive effects, there are also negative point estimates for specific players. For those players, the technical advantage might not be an actual advantage, even though this is not backed by statistical confidence. In total the individualized effects range from around –5 to +15 percentage points, suggesting that there is some heterogeneity in the effect.

A comparison of the 10% most affected (highest treatment effect) to the 10% least affected (lowest treatment effect) individuals in selected characteristics can be found in Table 3. This helps to evaluate the differences in specific characteristics for those with the highest and lowest treatment effects. The estimate represents the difference in the characteristics of the most and least affected. Joint p-values in contrast to the 'usual' p-values account for testing the estimates of, in this case, 10 tests (for more details the interested reader is referred to Chernozhukov, Fernández-Val, and Luo, 2018).

*Table 3: Difference in characteristics of 10% most and least affected groups*

| Characteristics | Estimate | Joint p-value | p-value |
|---|---|---|---|
| Ranking | 360.89 (109.66) | 0.05 | 0.00 |
| 3-darts average | -5.54 (1.79) | 0.07 | 0.00 |
| Acc. Matches | -488.31 (96.48) | 0.00 | 0.00 |
| Years playing | -13.13 (3.78) | 0.03 | 0.00 |
| Years prof. | -11.34 (2.38) | 0.00 | 0.00 |
| Age | -6.87 (4.05) | 0.66 | 0.05 |
| Venue in country of birth | 0.05 (0.21) | 1.00 | 0.40 |
| Home | -0.49 (0.13) | 0.02 | 0.00 |
| Prize money standardized | -0.28 (0.09) | 0.09 | 0.00 |
| Best of legs | -2.54 (2.62) | 0.98 | 0.17 |

Notes: Bias corrected, 999 (weighted) bootstrap replications. Joint p-values account for the simultaneous inference conducted for all (10) differences of variables. The estimate is the difference in the characteristics for the most affected (highest) minus the least affected (lowest) treatment effect).

In general, the most affected, i.e. those with the highest realization of the technical advantage, have lower performance measures, less experience, competing away from home, and in tournaments with lower prize money, compared to the least affected. A clearly insignificant estimate can be found for *age*, which is arguably the worst proxy available for experience. Furthermore, only *home* as defined as the hometown being within a radius of 100km to the tournament venue seems to be differently allocated, while being born in the country (*Venue in country of birth*) of the venue is almost equally frequent among most and least affected.

## 5.3  Group average treatment effects

Now we turn to investigating more specific effect heterogeneities. Two group effects are investigated: First, in Section 5.3.1, we analyse if there are differential effects in the BIA associated with the ability of the contestants. Second, we evaluate in Section 5.3.2 whether, motivated by the social pressure hypothesis, there are differential effects associated to competitors performing near their hometown.

**5.3.1 Ability**

A contest is fair if equally skilled contestants have equal winning probabilities. To discover if the BIA differs for equally skilled contestants on any level of ability, we therefore analysed the BIA of the first moving player (contestant *i*) associated with both competitors' ability (of *i* and *j*). As already discussed, our preferred measure of ability is the performance measure 3-darts average.

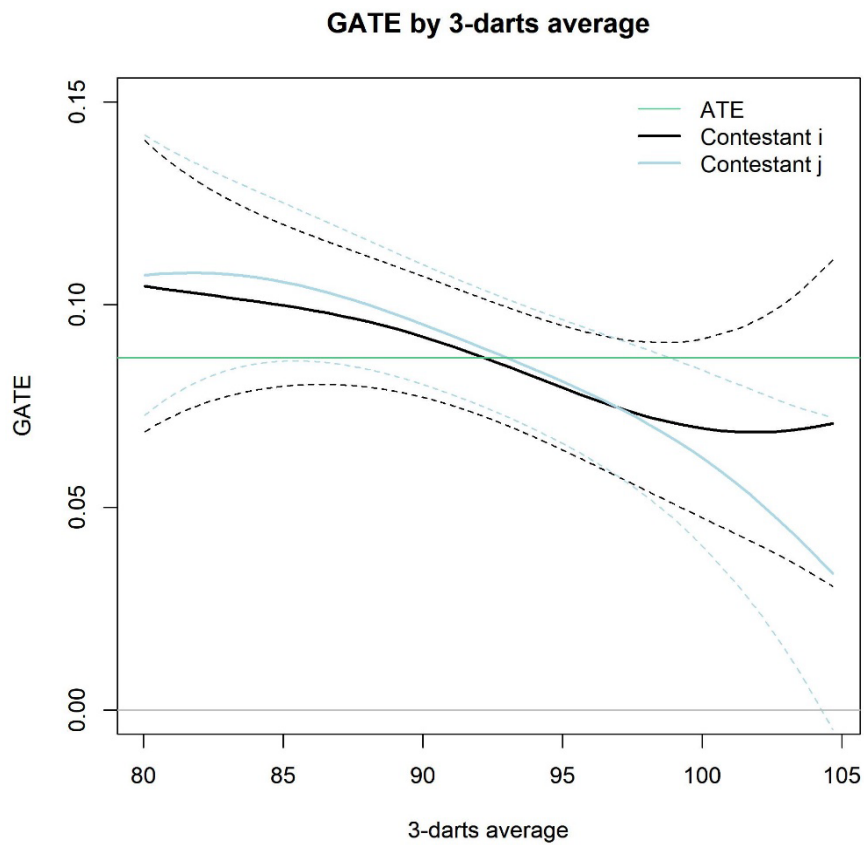

*Figure 2: Built-in advantage of starting contestant by ability*

Notes: The broken lines represent the 90% confidence intervals. Contestant *i* is the starting contestant, *j* the opponent.

In general, Figure 2 shows the GATEs to be smaller for higher ability athletes, in line with the results in Table 3. However, both contestants' GATEs are close to each other for every value of the 3-darts average. In other words, there is no evidence that the BIAs are different for equally skilled contestants. Note that, especially for low and high 3-darts average, there are fewer observations, resulting in larger confidence intervals.

### 5.3.2 Social incentives

We observe in Table 3 that those with the highest treatment effects are rather competing away from home, with an estimated difference of -0.49, compared to those with the lowest treatment effects. Investigating this more specifically, the group average effects are shown in Table 2. Column (2) presents the effect associated with competing at home (-0.0885) relative to those performing not at home. The technical advantage therefore, exists only for the group of individuals not performing at home (0.0939), and close to zero for those performing at home. The diminished winning probability of about 8.8 percentage points for the group of contestants performing in a friendly environment with support, i.e. at home, is in line with the social pressure and contrary to the social support hypothesis, as found in Dohmen (2008) and Harb-Wu and Krumer (2019).[19] No substantial difference is found for the group of competitors born in the same country as the venues' country (0.0032, column (3) in Table 2). This indicates that the differential effect associated with competing at home can be attributed to social pressure from the audience, while there is no difference in the effect associated with the audience cheering for their compatriots.

*Table 4: Social pressure vs. home (dis-)advantage*

|  | (1) ATE | (2) Home ($i$) | (3) Home ($j$) | (4) Home ($i$ & $j$) |
|---|---|---|---|---|
| Intercept | 0.0865*** | 0.0939*** | 0.0852*** | 0.0926*** |
|  | (0.0089) | (0.0093) | (0.0093) | (0.0097) |
| Home (Contestant $i$) |  | -0.0885*** |  | -0.0887*** |
|  |  | (0.0323) |  | (0.0323) |
| Home (Contestant $j$) |  |  | 0.0144 | 0.0168 |
|  |  |  | (0.0323) | (0.0323) |
| Observations | 11604 | 11604 | 11604 | 11604 |

Notes: Best linear prediction as described in Semenova and Chernozhukov (2017). Heteroscedasticity robust standard error in parentheses. *Home* is defined as the hometown of the contestant being in the radius of 100km to the venue. Contestant $i$ is the starting contestant, $j$ is the opponent. *, **, and *** represent statistical significance at the 10, 5, and 1% level, respectively.

---

[19] Changing the definition of home in terms of a different radius around the venue can be found in Table A.2 in Appendix B.3. For a narrower definition of *home* contestants, the associated effect is larger, while the associated effect decreases if defining contestants' farther away from the venue as *home* contestants. This supports the general conclusion outlined previously.

Interestingly, there is no differential effect for the second moving contestant (*j*) to be affected by social pressure or any type of home (dis-)advantage, as seen in Table 4. Columns (1) and (2) are equivalent to those presented in Table 2; columns (3) and (4) show the differential effect for non-starting contestants associated to competing in a venue near their hometown. In other words, competing in a friendly environment has no (negative) effect on the BIA for non-starting contestants. Therefore, there is no evidence for a general home (dis-)advantage, but evidence in line with social pressure for those starting the match.

On the contrary, an analogy with the finding of Harb-Wu and Krumer (2019), that the effect is driven by those athletes in the top quartile of the ability distribution, cannot be confirmed.

*Figure 3: Differential built-in advantage by home and ability*

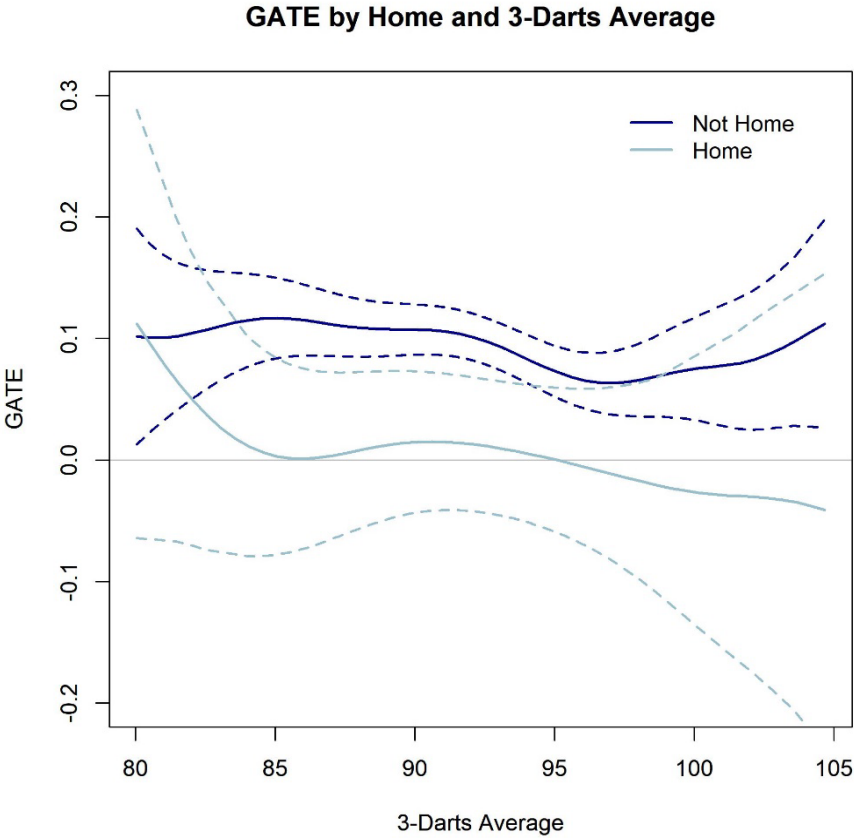

Notes: The dark blue line represents the GATEs for contestants not home, the light blue line for contestants performing at home. The broken lines represent the 90% confidence intervals.

Figure 3 displays the treatment effect for the starter associated with performing at home or not at home with respect to the 3-darts average, the most accurate measure of individual ability in darts. We discovered that the effect associated with playing at home fluctuates around the zero effect, while the effect for those playing away from home is always above. Especially for the boundaries of the figure, i.e. low and high 3-darts average, there are less observations; therefore, the estimates become imprecise. Replacing the 3-darts average by the previous position of the player in the ranking leads to similar conclusions; the results can be found in Figure A.2 in Appendix B.2.

## 5.4 Sensitivity checks

To assess the sensitivity of the results being method dependent we chose to additionally perform the analysis using a Modified Causal Forest (Lechner, 2018). For more details on the method and implementation the interested reader can refer to Appendix D.

*Table 5: Average and Group Average Effects, MCF*

| Effects | Estimate | Standard Error | t- value | p- value |
|---|---|---|---|---|
| ATE | 0.0838 | 0.0147 | 5.70 | 0.00 |
| GATE (home=0) | 0.0911 | 0.0151 | 6.03 | 0.00 |
| GATE (home=1) | 0.0038 | 0.0357 | 0.11 | 0.91 |
| Difference | 0.0873 | 0.0353 | 2.47 | 0.01 |
| With sample-based clustering | | | | |
| ATE | 0.0872 | 0.0152 | 5.74 | 0.00 |
| GATE (home=0) | 0.0931 | 0.0155 | 6.01 | 0.00 |
| GATE (home=1) | 0.0225 | 0.0365 | 0.62 | 0.54 |
| Difference | 0.0706 | 0.0363 | 1.94 | 0.05 |

Notes: Average Treatment Effect (ATE) and Group Average Treatment Effects (GATE), associated with performing in a venue near the hometown (home=1) and not near the hometown (home=0). *Home* is defined as performing at a tournament venue that is within a radius of 100km to the hometown of the athlete. Estimated using a Modified Causal Forest. Inference is conducted as weight-based inference. In the lower part of the table, effects are estimated using sample-based clustering on the individual player level; 1000 Trees with 50% subsampling rate and the minimum leave size in each tree equals 2.

All estimated effects are in a comparable range to the estimates in the previous sections. In addition, the conclusions drawn hold in general. Table 5 shows the ATE estimate, as well as the GATE associated with performing at home resulting from the Modified Causal Forest estimation. Further, in the lower part of the table, we repeated the estimation but accounted for contestant specific clusters. For both estimations, we found a large differential effect associated to performing at home.

For the ability measure, 3-darts average, the pattern is in line with the findings from Section 5.3.1. In Figure A.3 in Appendix B.4 we observe lower GATEs for higher 3-darts averages of the starting contestant (i) and the opponent (j). Especially, for equal levels of ability the GATEs are similar. We can therefore conclude that the BIA is symmetric for equally skilled contestants. Note that, since the 3-darts average is discretized to form groups of roughly equal numbers of observation in each group, we can see that especially for low and high 3-darts averages there are less observations. In conclusion, this conceptually different method produces similar results, providing some evidence that the results presented in the previous sections are robust.

Table 6: Base results - best linear predictors, sensitivity to clustering

|  | (1) ATE | (2) Home | (3) Country of birth | (4) All |
|---|---|---|---|---|
| Intercept | 0.0875*** | 0.0948*** | 0.0863*** | -0.0104 |
|  | (0.0095) | (0.0100) | (0.0120) | (0.2316) |
| Home |  | -0.0878*** |  | -0.0968*** |
|  |  | (0.0338) |  | (0.0359) |
| Venue in country of birth |  |  | 0.0030 | 0.0238 |
|  |  |  | (0.0214) | (0.0225) |
| Prize money standardized |  |  |  | -0.0936* |
|  |  |  |  | (0.0488) |
| Ranking |  |  |  | 0.0000 |
|  |  |  |  | (0.0001) |
| 3-darts average |  |  |  | 0.0011 |
|  |  |  |  | (0.0025) |
| Acc. matches |  |  |  | -0.0001 |
|  |  |  |  | (0.0001) |
| Years playing |  |  |  | -0.0016 |
|  |  |  |  | (0.0016) |
| Years prof. |  |  |  | -0.0024 |
|  |  |  |  | (0.0019) |
| Age |  |  |  | 0.0019 |
|  |  |  |  | (0.0016) |
| Best of legs |  |  |  | 0.0011 |
|  |  |  |  | (0.0018) |
| Observations | 11604 | 11604 | 11604 | 11604 |

Notes: Best linear prediction as described in Semenova and Chernozhukov (2017); cluster-based sampling and clustered standard errors on the individual level in parentheses. *Home* is defined as the hometown of the contestant being in the radius of 100km to the venue. *, **, and *** represent statistical significance at the 10, 5, and 1% level, respectively.

Accounting for potential contestant specific clusters, the best linear prediction results from Table 2 are replicated with cluster-based sampling for the nuisance estimation, as well as clustered standard errors on the individual level in Table 6. While standard errors are slightly higher, we find the conclusions drawn previously are valid.

## 5.5  Discussion

A contest design in which one group systematically realizes differential BIAs is unfair. Especially when home contestants are affected, the competition design must be reconsidered and changed to have more fair and interesting contests.

*Figure 4: BIA by maximum number of legs*

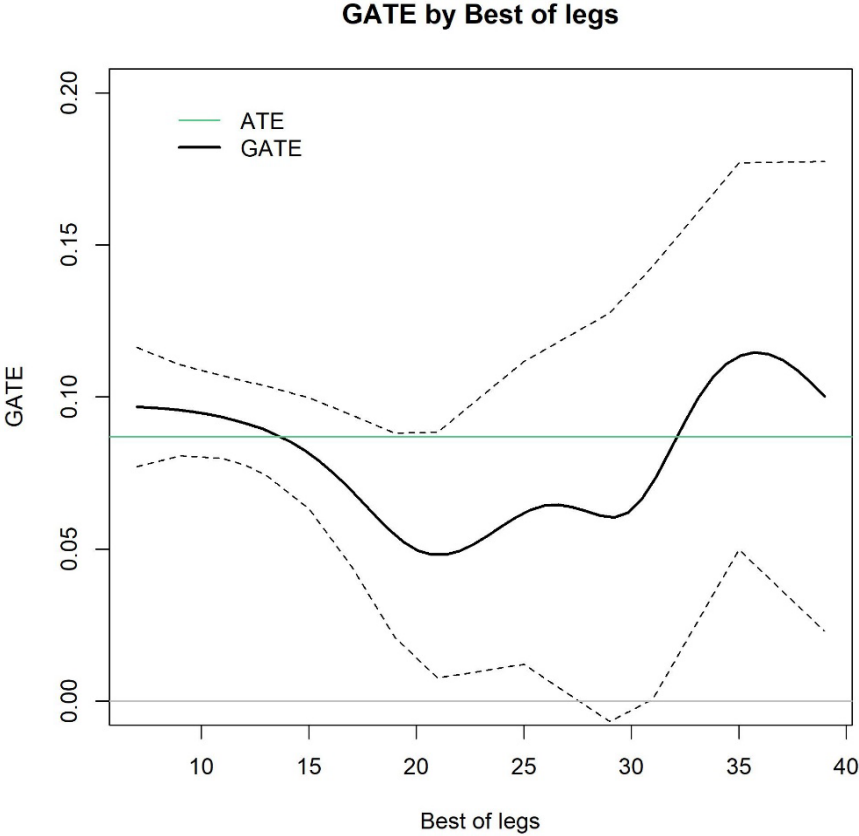

Notes: The GATE is shown as solid black line; the broken lines represent the 90% confidence intervals. The green line represents the ATE.

A potential idea is to increase the length of the contest, such that the BIA decreases. While that looks like a neat idea in theory, we cannot provide clear evidence for it. Figure 4 provides the BIA associated to the number of maximum legs to play in a match.

The GATE fluctuates around the average effect, with no clear tendency towards lower or no BIA for longer matches. A second option is to remove the technical advantage from the contest design in form of even numbers of maximum legs to play, as for example, in tennis tournaments.[20] Further, Cohen-Zada, Krumer, and Shapir (2018) found that the order of serves in tennis tiebreak, which is the ABBA sequence, does not provide any advantage to any of the players. Adopting a similar structure might help to improve the fairness in darts contests.

# 6 Conclusion

The study discovered an average effect of about 8.6 percentage points higher winning probability induced by the BIA in the sequential darts contest with two players. Having a randomized allocation of being the first-mover for equally skilled contestants, the contest is fair for contestants with symmetric potential BIAs. While we found non-differential BIAs for equally skilled contestants, the personalized point estimates suggested substantial heterogeneity in the effect, ranging from about -5 to +15 percentage points with athletes having more experience and better performance measures to benefit least. In addition, estimated differential effects were in line with the social pressure hypothesis as found in previous works, especially in the behavioural economics literature. For the group of individuals competing in a friendly environment, the BIA was lower compared to those competing on neutral grounds, which was in line with the works of Dohmen (2008) and Harb-Wu and Krumer (2019) among others. This leads to not necessarily equal winning probabilities for contestants with equal abilities, and therefore, to an unfair contest.

For designing contests, the heterogeneous nature of the BIA in such types of contests should be considered to not unintentionally favour specific groups of contestants. Especially if fairness is a precondition for a contest, the presented findings should be taken into account. Despite fairness issues, local athletes might be important for the attractiveness of the contest. Introducing a fairer contest design may, therefore, improve the attractiveness as well. We did not find any evidence that an increase in the maximum length for a match leads to diminished BIAs, so the suggestion would be to change the design of the contest, such that there is no BIA for any contestant. Furthermore, the findings are of interest to individuals performing in situations of increased importance, either contestants in tournaments or in daily life situations. While preparing for those situations, one should put more emphasis on mental training.

---

[20] Another potential idea to resolve the BIA is to give the respective second-mover the chance to catch-up in case the starter of the leg finishes the leg.

Another insight from this study is that *Causal Machine Learning* is highly valuable for empirical studies. It enables us to provide a more complete picture of particular effects under investigation, as well as a way to investigate specific hypotheses in a flexible way. Furthermore, the two steps of removing potential selection bias and estimating treatment effects offers an intuitive, flexible, and robust way to improve credibility of empirical research.

Finally, we call for more, fine granular investigations in contest designs, particularly in situations in which there are potentially multiple incentives observable (and distinguishable). Since we only have data on male professional players, we encourage further studies to analyse how the BIA affects women, as well as youth and amateur players in darts or other asymmetric contests. Furthermore, in future studies, the limitations of audience presence owing to the current corona pandemic could be used to analyse the BIA, without the direct influence coming from the audience. Methods from the *Causal Machine Learning* toolbox appear to be especially useful for this and should be considered for further empirical analyses.

# References

Apesteguia, J., & Palacios-Huerta, I. (2010). Psychological pressure in competitive environments: Evidence from a randomized natural experiment. *American Economic Review, 100*(5), 2548-2564.
Ariely, D., Gneezy, U., Loewenstein, G., & Mazar, N. (2009). Large Stakes and Big Mistakes. *The Review of Economic Studies, 76*(2), 451-469.

Arlegi, R., & Dimitrov, D. (2020). Fair elimination-type competitions. *European Journal of Operational Research, 287*(2), 528-535.

Athey, S. (2017). Beyond prediction: Using big data for policy problems. *Science, 355*, 483-485.

Athey, S., & Imbens G. (2019). Machine Learning Methods Economists Should Know About. *arXiv:1903.10075*.

Athey, S., & Imbens, G. (2016). Recursive partitioning for heterogeneous causal effects. *Proceedings of the National Academy of Sciences, 113*(27), 7353-7360.

Athey, S., & Wager, S. (2019). Estimating Treatment Effects with Causal Forests: An Application. *arXiv:1902.07409*.

Athey, S., Tibshirani, J., & Wager, S. (2019). Generalized random forests. *Annals of Statistics, 47*(2), 1179-1203.

Baumeister, R. (1984). Choking under pressure: Self-consciousness and paradoxical effects of incentives on skillful performance. *Journal of Personality and Social Psychology, 46*(3), 610-620.

Baumeister, R., Hamilton, J., & Tice, D. (1985). Public Versus Private Expectancy of Success: Confidence Booster or Performance Pressure? *Journal of Personality and Social Psychology, 48*(6), 1447-1457.

Breiman, L. (2001). Random Forests. *Machine Learning, 45*, 5-32.

Butler, J., & Baumeister, R. (1998). The trouble with friendly faces: Skilled performance with a supportive audience. *Journal of personality and social psychology, 75*(5), 1213.

Cao, Z., Price, J., & Stone, D. (2011). Performance under pressure in the NBA. *Journal of Sports Economics, 12*(3), 231-252.

Chernozhukov, V., Chetverikov, D., Demirer, M., Duflo, E., Hansen, C., Newey, W., & Robins, J. (2018). Double/debiased machine learning for treatment and structural parameters. *The Econometrics Journal, 21*(1), C1-C68.

Chernozhukov, V., Fernández-Val, I., & Luo, Y. (2018). The Sorted Effects Method: Discovering Heterogeneous Effects Beyond Their Averages. *Econometrica, 86*(6), 1911-1938.

Cockx, B., Lechner, M., & Bollens, J. (2019). Priority to unemployed immigrants? A causal machine learning evaluation of training in Belgium. *arXiv: 1912.12864*.

Cohen-Zada, D., Krumer, A., & Shapir, O. (2018). Testing the effect of serve order in tennis tiebreak. *Journal of Economic Behavior and Organization, 146*, 106-115.

Davis, J., & Heller, S. (2017). Using causal forests to predict treatment heterogeneity: An application to summer jobs. *American Economic Review: Papers & Proceedings, 107*(5), 546-550.

Dohmen, T. (2008). Do professionals choke under pressure? *Journal of Economic Behavior and Organization, 65*, 636-653.

Ehrenberg, R., & Bognanno, M. (1990). Do tournaments have incentive effects? *Journal of Political Economy, 98*(6), 1307-1324.

Faltings, R., Krumer, A., & Lechner, M. (2019). Rot-Jaune-Verde. Language and Favoritism: Evidence from Swiss Soccer. *Economic Working Paper Series, 1915*.

Fan, Q., Hsu, Y.-C., Lieli, R., & Zhang, Y. (2019). Estimation of Conditional Average Treatment Effects with High-Dimensional Data. *arXiv:1908.02399*.

Ginsburgh, V., & Van Ours, J. (2003). Expert opinion and compensation: Evidence from a musical competition. *American Economic Review, 93*(1), 289-296.

Goller, D., & Krumer, A. (2020). Let's meet as usual: Do games played on non-frequent days differ? Evidence from top European soccer leagues. *European Journal of Operational Research, 286*(2), 740-754.

Goller, D., Knaus, M., Lechner, M., & Okasa, G. (2018). Predicting Match Outcomes in Football by an Ordered Forest Estimator. *Economic Working Paper Series, No. 1811*.

González-Díaz, J., Gossner, O., & Rogers, B. (2012). Performing best when it matters most: Evidence from professional tennis. *Journal of Economic Behavior and Organization, 84*(3), 767-781.

Harb-Wu, K., & Krumer, A. (2019). Choking Under Pressure in Front of a Supportive Audience: Evidence from Professional Biathlon. *Journal of Economic Behavior and Organization, 166*, 246-262.

Hastie, T., Tibshirani, R., & Friedman, J. (2009). *The Elements of Statistical Learning - Data mining, inference, and prediction.* 2nd. ed. New York: Springer.

Holland, P. (1986). Statistics and causal inference. *Journal of the American Statistical Association, 81*(396), 945-960.

Hurley, W. (2009). Equitable birthdate categorization systems for organized minor sports competition. *European Journal of Operational Research, 192*(1), 253-264.

Kirkegaard, R. (2012). Favoritism in asymmetric contests: Head starts and handicaps. *Games and Economic Behavior, 76*(1), 226-248.

Klein Teeselink, B., Potter Van Loon, R., Van Den Assem, M., & Van Dolder, D. (2020). Incentives, Performance and Choking in Darts. *Journal of Economic Behavior and Organization, 169*, 38-52.

Knaus, M. (2020). Double Machine Learning based Program Evaluation under Unconfoundedness. *arXiv:2003.03191v1*.

Knaus, M., Lechner, M., & Strittmatter, A. (2020a). Machine Learning Estimation of Heterogeneous Causal Effects: Empirical Monte Carlo Evidence. *The Econometrics Journal, utaa014*.

Knaus, M., Lechner, M., & Strittmatter, A. (2020b). Heterogeneous Employment Effects of Job Search Programmes: A Machine Learning Approach. *Journal of Human Resources, 0718-9615R1*.

Konrad, K. (2002). Investment in the absence of property rights; the role of incumbency advantages. *European Economic Review, 46*, 1521-1537.

Krumer, A., & Lechner, M. (2018). Midweek effect on soccer performance: Evidence from the German Bundesliga. *Economic Inquiry, 56*(1), 193-207.

Lazear, E. (2000). The Power of Incentives. *The American Economic Review: Papers and Proceedings, 90*(2), 410-414.

Lechner, M. (2018). Modified Causal Forests for Estimating Heterogeneous Causal Effects. *arXiv:1812.09487v2*.

Levitt, S., & List, J. (2008). Homo economicus evolves. *Science, 319*(5865), 909-910.

Liebscher, S., & Kirschstein, T. (2017). Predicting the outcome of professional darts tournaments. *International Journal of Performance Analysis in Sport, 17*(5), 666-683.

Masters, R. (1992). Knowledge, knerves and know-how: The role of explicit versus implicit knowledge in the breakdown of a complex motor skill under pressure. *British Journal of Psychology, 83*(3), 343-358.

Meirowitz, A. (2008). Electoral contests, incumbency advantages, and campaign finance. *Journal of Politics, 70*(3), 681-699.

Musch, J., & Grondin, S. (2001). Unequal competition as an impediment to personal development: A review of the relative age effect in sport. *Developmental Review, 21*(2), 147-167.

Nie, X., & Wager, S. (2017). Quasi-Oracle Estimation of Heterogeneous Treatment Effects. *arXiv:1712.04912v3*.

Nitzan, S. (1994). Modelling rent-seeking. *European Journal of Political Economy, 10*, 41-60.

Ötting, M., Deutscher, C., Schneemann, S., Langrock, R., Gehrmann, S., & Scholten, H. (2020). Performance under pressure in skill tasks: An analysis of professional darts. *PLoS ONE, 15*(2), 1-21.

Page, L., & Page, K. (2007). The second leg home advantage: Evidence from European football cup competitions. *Journal of Sports Sciences, 25*(14), 1547-1556.

Prendergast, C. (1999). The Provision of Incentives in Firms. *Journal of Economic Literature, 37*(1), 7-63.

Robins, J., Rotnitzky, A., & Zhao, L. (1994). Estimation of Regression Coefficients When Some Regressors Are Not Always Observed. *Journal of the American Statistical Association, 89*(427), 846-866.

Robins, J., Rotnitzky, A., & Zhao, L. (1995). Analysis of Semiparametric Regression Models for Repeated Outcomes in the Presence of Missing Data. *Journal of the American Statistical Association, 90*(429), 106-121.

Rosen, S. (1986). Prizes and Incentives in Elimination Tournaments. *The American Economic Review, 76*(4), 701-715.

Rubin, D. (1974). Estimating causal effects of treatments in randomized and nonrandomized studies. *Journal of Educational Psychology, 66*(5), 688-701.

Segev, E., & Sela, A. (2014). Sequential all-pay auctions with head starts. *Social Choice and Welfare, 43*(4), 893-923.

Semenova, V., & Chernozhukov, V. (2017). Simultaneous Inference for Best Linear Predictor of the Conditional Average Treatment Effect and Other Structural Functions. *arXiv:1702.06240v3*.

Shapiro, C., & Stiglitz, J. (1984). Equilibrium Unemployment as a Worker Discipline Device. *The American Economic Review, 74*(3), 433-444.

Stiglitz, J. (1976). The Efficiency Wage Hypothesis, Surplus Labour, and the Distribution of Income in LDCs. *Oxford Economic Papers, 28*(2), 185-207.

Strauss, B. (1997). Choking under pressure: Positive public expectations and performance in a motor task. *Zeitschrift für Experimentelle Psychologie, 44*(4), 636-655.

Szymanski, S. (2003). The Economic Design of Sporting Contests. *Journal of Economic Literature, XLI*, 1137-1187.

Tian, L., Alizadeh, A., Gentles, A., & Tibshirani, R. (2014). A simple method for estimating interactions between a treatment and a large number of covariates. *Journal of the American Statistical Association, 109*(508), 1517-1532.

Tibshirani, R., Price, A., & Taylor, J. (2011). A statistician plays darts. *Journal of the Royal Statistical Society. Series A: Statistics in Society, 174*(1), 213-226.

Toma, M. (2017). Missed Shots at the Free-Throw Line: Analyzing the Determinants of Choking Under Pressure. *Journal of Sports Economics, 18*(6), 539-559.

Tullock, G. (1980). Efficient rent-seeking. *In: J.M. Buchanan, R.D. Tollison and G. Tullock (Eds.), Toward a theory of the rent-seeking society. College Station: Texas A&M University Press*, 97-112.

Wager, S., & Athey, S. (2018). Estimation and Inference of Heterogeneous Treatment Effects using Random Forests. *Journal of the American Statistical Association, 113*(523), 1228-1242.

Wager, S., & Walther, G. (2015). Adaptive Concentration of Regression Trees, with Application to Random Forests. *arXiv:1503.06388*.

Wright, M. (2014). OR analysis of sporting rules - A survey. *European Journal of Operational Research, 232*(1), 1-8.

Zajonc, R. (1965). Social Facilitation. *Science, 149*(3681), 269-274.

Zimmert, M., & Lechner, M. (2019). Nonparametric estimation of causal heterogeneity under high-dimensional confounding. *arXiv:1908.08779*.

# Appendices

## Appendix A: Descriptive statistics

*Table A.1: Descriptive statistics, all variables*

| Variable | Mean | Starting Player | Non-Starting Player |
|---|---:|---:|---:|
| Game Outcomes | | | |
| Wins Match | 0.50 | 0.55 | 0.45 |
| Wins First Leg | 0.50 | 0.63 | 0.37 |
| Game Characteristics | | | |
| Starts First Leg | 0.50 | 1.00 | 0.00 |
| Best of Legs | 11.21 (3.43) | | |
| Televised | 0.32 | | |
| Ranking Tournament | 0.90 | | |
| BDO Tournament | 0.04 | | |
| Double In | 0.01 | | |
| Prize Money | 51145 (79918) | | |
| Prize Money (standardized) | 0.13 (0.25) | | |
| Player Characteristics | | | |
| 3 Darts Average | 91.82 (4.53) | 91.99 (4.49) | 91.65 (4.56) |
| Ranking Last Year | 121.74 (292.77) | 113.00 (254.99) | 130.47 (326.20) |
| Accumulated Matches | 231.15 (270.17) | 238.13 (274.46) | 224.17 (265.81) |
| Left handed | 0.07 | 0.07 | 0.07 |

*Table A.1 continued*

| Variable | Mean | Starting Player | Non-Starting Player |
|---|---:|---:|---:|
| Years Playing | 19.97 (9.82) | 20.07 (9.79) | 19.87 (9.85) |
| Years Playing Professional | 11.91 (6.21) | 11.90 (6.17) | 11.91 (6.25) |
| Age | 36.54 (9.73) | 36.46 (9.63) | 36.61 (9.83) |
| Venue in country of birth | 0.38 | 0.38 | 0.38 |
| Home (50km radius) | 0.04 | 0.03 | 0.04 |
| Home (100km radius) | 0.08 | 0.08 | 0.08 |
| Home (150km radius) | 0.15 | 0.15 | 0.15 |
| Home (200km radius) | 0.21 | 0.22 | 0.21 |
| Distance to Venue | 1410.25 (3271.76) | 1422.84 (3307.26) | 1397.66 (3241.93) |
| Diff. 3 Darts Average | | 0.34 (5.76) | -0.34 (5.76) |
| Diff. Ranking last year | | -17.46 (406.21) | 17.46 (406.21) |
| Diff. Accumulated Matches | | 13.96 (355.44) | -13.96 (355.44) |
| Diff. Years Playing | | 0.20 (13.89) | -0.20 (13.89) |
| Diff. Years Playing Professional | | -0.01 (8.76) | 0.01 (8.76) |
| Diff. Age | | -0.15 (13.76) | 0.15 (13.76) |
| Diff. Distance to Venue | | 25.18 (4309.09) | -25.18 (4309.09) |
| Year | | | |
| 2009 | 0.01 | | |
| 2010 | 0.02 | | |
| 2011 | 0.02 | | |
| 2012 | 0.02 | | |
| 2013 | 0.02 | | |
| 2014 | 0.02 | | |
| 2015 | 0.03 | | |
| 2016 | 0.07 | | |
| 2017 | 0.07 | | |
| 2018 | 0.34 | | |
| 2019 | 0.37 | | |
| Observations | 11604 | 11604 | 11604 |

Notes: This table presents average values and standard deviations (in parentheses for non-binary variables). Values that are equal for all columns are reported only once.

# Appendix B: Additional results

## Appendix B.1: Common support assumption

*Figure A.1: Propensity scores*

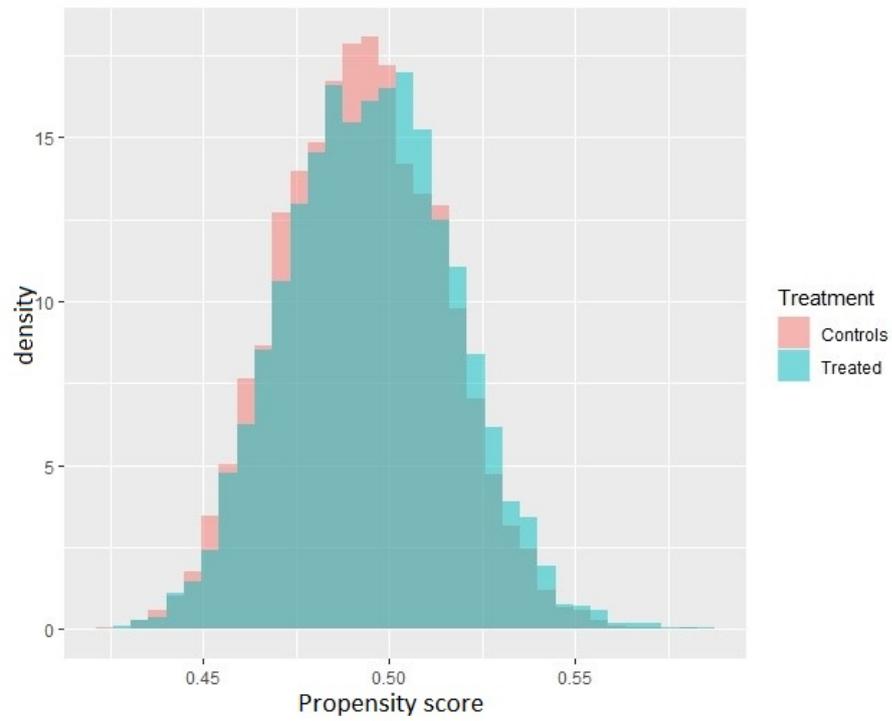

Notes: Shown are the propensity scores for controls (in red) and treated (in blue). On the horizontal line you find the value of the propensity score

# Appendix B.2: Alternative ability measure

*Figure A.2: Differential built-in advantage by ability, alternative specification*

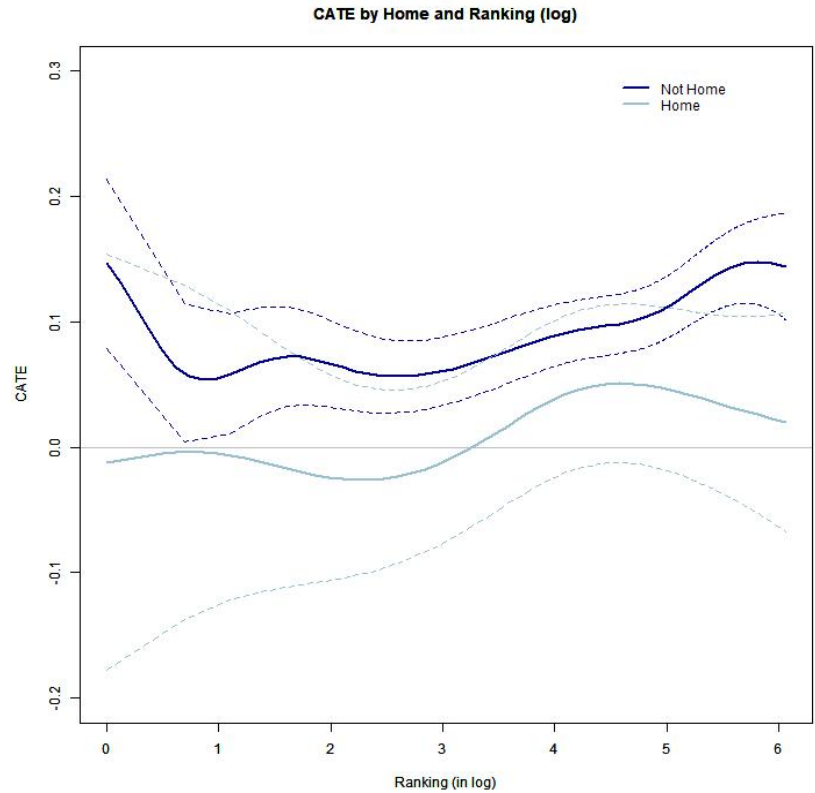

Notes: The dark blue line represents the GATEs for contestants not home, the light blue line for contestants performing at home. The broken lines represent the 90% confidence intervals.

# Appendix B.3: Alternative definitions of *home*

*Table A.2: Different definitions of 'Home'*

|          | (1) 50 km    | (2) 100 km   | (3) 150 km   | (4) 200 km  | (5) Distance (in km) |
|----------|--------------|--------------|--------------|-------------|----------------------|
| Constant | 0.0921***    | 0.0939***    | 0.0962***    | 0.0947***   | 0.0877***            |
|          | (0.0091)     | (0.0093)     | (0.0097)     | (0.0100)    | (0.0097)             |
| Home     | -0.1533***   | -0.0885***   | -0.0664***   | -0.0393*    | -0.0000              |
|          | (0.0471)     | (0.0323)     | (0.0249)     | (0.0217)    | (0.0000)             |

Notes: Best linear prediction as described in Semenova and Chernozhukov (2017). Heteroscedasticity robust standard error in parentheses. *, **, and *** represent statistical significance at the 10, 5, and 1% level, respectively. Different columns represent different definitions of the *Home* variable, for which the hometown of the contestant is within 50 km (column 1), 100 km (2), 150km (3), 200km (4), and the continuous measure for the distance in km (5), to the venue.

## Appendix B.4: Sensitivity checks

*Figure A.3: Built-in advantage of starting contestant by ability; MCF*

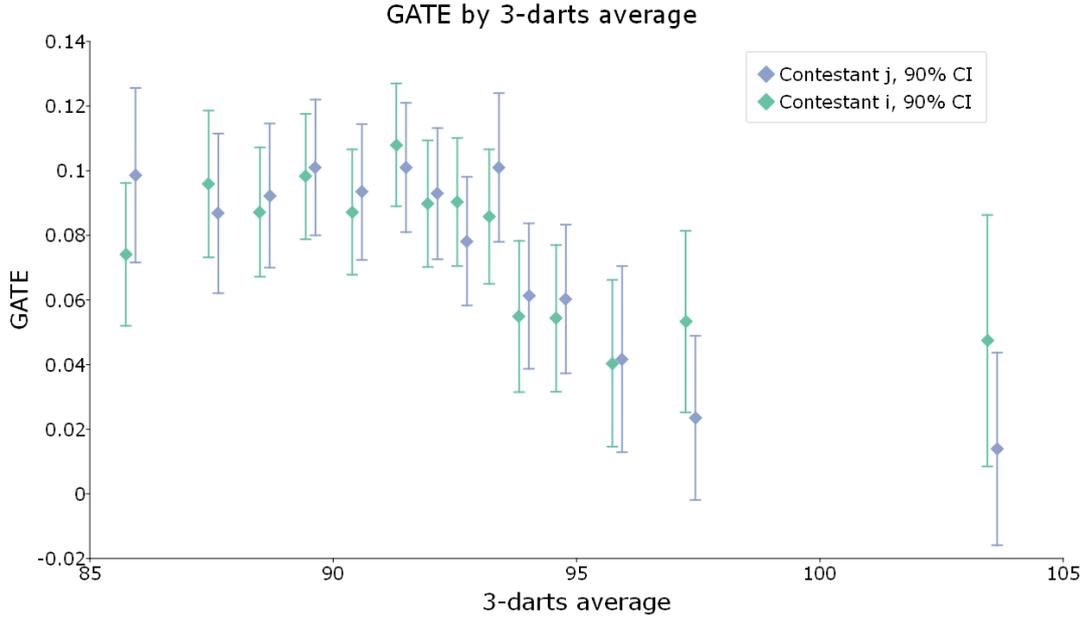

Notes: Contestant *i* is the starting player depicted in blue, accompanied by the 90% confidence intervals.
Contestant *j* is the opponent depicted in green. Standard errors are calculated as weight-based standard errors.

## Appendix C: Contest model

We model a match in darts by adapting the lottery contest structure from Tullock (1980) contests, which is the major workhorse for modelling contests (for a survey on ways of modelling contests, see e.g., Nitzan, 1994) . The probability to win for contestant *i* starting is: $Pr_{i,Starter}^{Win} = \frac{A_i}{A_i + \delta_i A_j}$ and for contestant *j* being the starter: $Pr_{j,Starter}^{Win} = \frac{A_j}{A_j + \delta_j A_i}$. We denote the ability of contestants *i* and *j* by $A_i$ and $A_j$. A potential BIA for the contestants is denoted by $0 < \delta_i, \delta_j \leq 1$.[21] Symmetrically, following $Pr_{j,Non-Starter}^{Win} = 1 - Pr_{i,Starter}^{Win}$ and vice versa for contestant *i*, we get: $Pr_{i,Non-Starter}^{Win} = \frac{A_i}{A_i + 1/\delta_j A_j}$ and $Pr_{j,Non-starter}^{Win} = \frac{A_j}{A_j + 1/\delta_i A_i}$. With the shootout before a match, there is a mechanism to allocate the right to start the contest, denoted by $\pi(A_i, A_j)$ depending on the ability of contestants *i* and *j*. The ex-ante probability to win the contest for *i* is:

$$Pr_i^{Win} = \pi(A_i, A_j) \times \frac{A_i}{A_i + \delta_i A_j} + (1 - \pi(A_i, A_j)) \times \frac{A_i}{A_i + 1/\delta_j A_j} \qquad (1)$$

$$= \frac{A_i A_j(1+\delta_i\delta_j)+2\delta_j A_i^2}{2(A_i+\delta_i A_j)(\delta_j A_i + A_j)} + (1 - 2\pi(A_i, A_j)) \times \frac{A_i A_j(\delta_i\delta_j - 1)}{2(A_i+\delta_i A_j)(\delta_j A_i + A_j)} \qquad (2)$$

$$= \frac{(1+\delta_i\delta_j)+2\delta_j}{2(1+\delta_i)(\delta_j+1)} + (1 - 2\pi) \times \frac{(\delta_i\delta_j - 1)}{2(1+\delta_i)(\delta_j+1)} \qquad (3)$$

---

[21] If $\delta_i = \delta_j = 1$, there is no BIA in the contest. In contrast to the 'classical' Tullock lottery contest the presented model is more flexible. We neither assume that there is no BIA nor that they are necessarily equal for both contestants.

A contest is regarded as fair if contestants with equal abilities have equal winning probabilities. Therefore, from equation (2) to (3) contestants with equal ability, i.e. $A_i = A_j = 1$, are assumed.[22] From equation (3), it can be concluded that the contest provides equal winning probabilities for equally skilled contestants, i.e. 50 % each, in one of the two cases: 1.) there is no BIA ($\delta_i = \delta_j = 1$), or 2.), the treatment allocation is randomized ($\pi = 0.5$), and the BIA is equal for both contestants ($\delta_i = \delta_j$). The first case of no BIA can be tested empirically. For the second case, while treatment allocation can be assumed to be randomized for equally skilled contestants, the size and nature of the BIA are subject to an empirical analysis, which is the focus of Section 5.

## Appendix D: Modified causal forest

In addition to the DML procedure, we used a conceptually different method from the class of estimators, which also achieved good overall performance in the study of Knaus et al. (2020a), a variant of Causal Forests. More specifically, we employed the *Modified Causal Forest* (MCF) proposed in Lechner (2018).

The MCF offers a single procedure to estimate the various levels of aggregations, IATEs, GATEs, and ATE, coming with a unified inference approach. Further, it has an easy way of implementing sample-based clustering, which can be applied to account for contestant-specific clustering.

---

[22] Note that, for ease of exposition, for contestants with equal abilities $\pi(A_i, A_j) = \pi$.

Generally, a causal forest estimator is built as an ensemble of many different *Causal Trees*, going back to Athey and Imbens (2016). Each *Causal Tree* splits the sample according to specific values of covariates sequentially into finer strata, so called leaves, until some stopping criterion is reached. The splits are conducted in a way to maximize the variance of treatment effects. This, in turn, mitigates selection effects and potentially uncovers effect heterogeneity in one step. Within final leaves, treatment effects are computed as the difference of the mean of the outcomes of treated and control units. Building many such trees on subsampled data and taking the average over the trees' effect estimates are the basis of causal forest estimators (Wager and Athey, 2018; Athey, Tibshirani, and Wager, 2019). Lechner (2018) refined this algorithm by implementing an improved splitting rule.

A potential drawback of this approach is a possible efficiency loss as the weight-based inference approach requires sample splitting. Moreover, in comparison to the Nonparametric CATE estimation discussed in the main part of this work, variables used in the analysis of effect heterogeneity needs to be discrete. For further details on the estimator, the reader can refer to Lechner (2018).

## Appendix E: Tournaments

*Table A.3: List of tournaments*

| Tournaments | Ranking Tournament* | Federation |
| --- | --- | --- |
| BDO World Darts Championship | yes | BDO |
| BDO World Masters | yes | BDO |
| BDO World Trophy | yes | BDO |
| Champions League of Darts | no | PDC |
| European Championship | yes | PDC |
| European Tour | yes | PDC |
| Finder / Zuiderduin Masters | yes | BDO |
| Grand Slam of Darts | no (2009-2014) yes (2015-2019) | PDC / BDO |
| PDC World Darts Championship | yes | PDC |
| Premier League Darts | no | PDC |
| Players Championship | yes | PDC |
| The Masters | no | PDC |
| UK Open | yes | PDC |
| World Cup of Darts | no | PDC |
| World Grand Prix | yes | PDC |
| World Matchplay | yes | PDC |
| World Series of Darts | no | PDC |

Notes: *Ranking Tournaments are open for qualification. Non-Ranking Tournaments are invite-only tournaments. Money won in ranking tournaments count as *Order of Merit* ranking points. For more details, see: https://www.pdc.tv/pdc-order-merit-rules (PDC) and http://www.bdodarts.com/images/bdo-content/doc-lib/F/bdo-invitation-rules.pdf (BDO).

## Appendix F: List of sources

Dartsforwindows (software; from www.dartsforwin.com)

www.pdc.tv

www.bdodarts.com

www.dartsdatabase.co.uk

www.wikipedia.org

www.mastercaller.com/players

www.dartswdf.com/rules/